\documentclass[a4paper,reprint,aps,pre]{revtex4-2}
\usepackage{graphicx}
\usepackage{dcolumn}
\usepackage{bm}
\usepackage{color}
\usepackage{xcolor}
\usepackage{amssymb}

\begin{document}
\title{Stochastic resonance in a model of a periodically driven DNA : Multiple transitions, scaling and sequence dependence}
\author{Ramu Kumar Yadav${^1}$}
\email{ramukumar@iisermohali.ac.in}
\author{M. Suman Kalyan$^{1,2}$}
\email{maroju.sk@gmail.com}
\author{Rajeev Kapri${^1}$}
\email{rkapri@iisermohali.ac.in}
\author{Abhishek Chaudhuri${^1}$}
\email{abhishek@iisermohali.ac.in}
\affiliation{\small \it ${^1}$Department of Physical Sciences, Indian Institute of Science Education and Research Mohali, Sector 81, Knowledge City, S. A. S. Nagar, Manauli PO 140306, India}
\affiliation{\small \it ${^2}$Department of Physics, Institute of Aeronautical Engineering, Dundigal, Hyderabad - 500043, Telangana, India}
\date{\today}

\newcommand{\figOne}{
\begin{figure*}
    \includegraphics[width=0.75\textwidth]{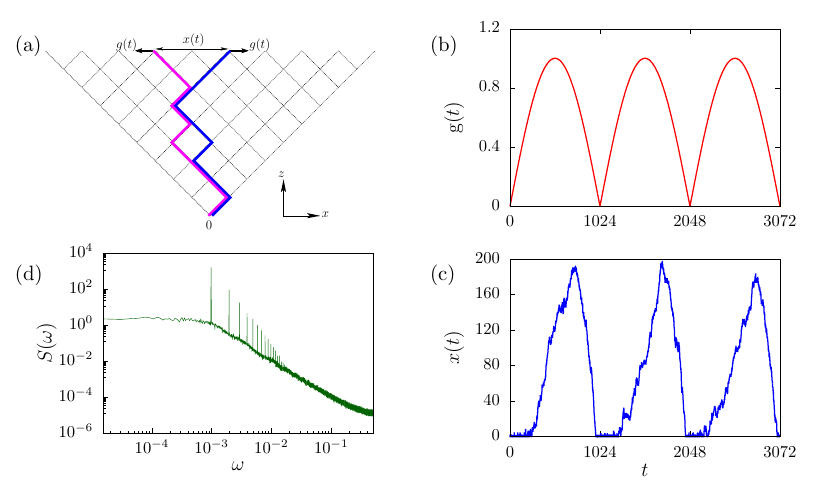}
	
	\caption{(a) Schematic diagram of the model. The strands of the DNA
	are shown by thick solid lines. One end of the DNA is anchored at
	the origin ($O$) and the free end monomers of the strands are pulled
	along the $x$ direction with the periodic force $g(t)$. (b) $g(t)$
	as a function of time $t$ for three consecutive cycles at a
	frequency $\omega_0 = 3.06 \times 10^{-3}$ and amplitude $G = 1.0$.
	(c) Separation between the end monomers, $x(t)$, follows the
	external force $g(t)$ with a lag. Here, length of the DNA $N = 128$.
	(d) Power spectral density $S(\omega)$ of the extension $x(t)$
	showing a background spectral density and delta peaks. $\eta$ in Eq.
	(3) is calculated from $S(\omega)$ using $\Delta \omega \approx 1.53
	\times 10^{-5}$.  }\label{fig:1}

\end{figure*}
}

\newcommand{\figFour}{
	\begin{figure}[t]
		\includegraphics[width=0.5\textwidth]{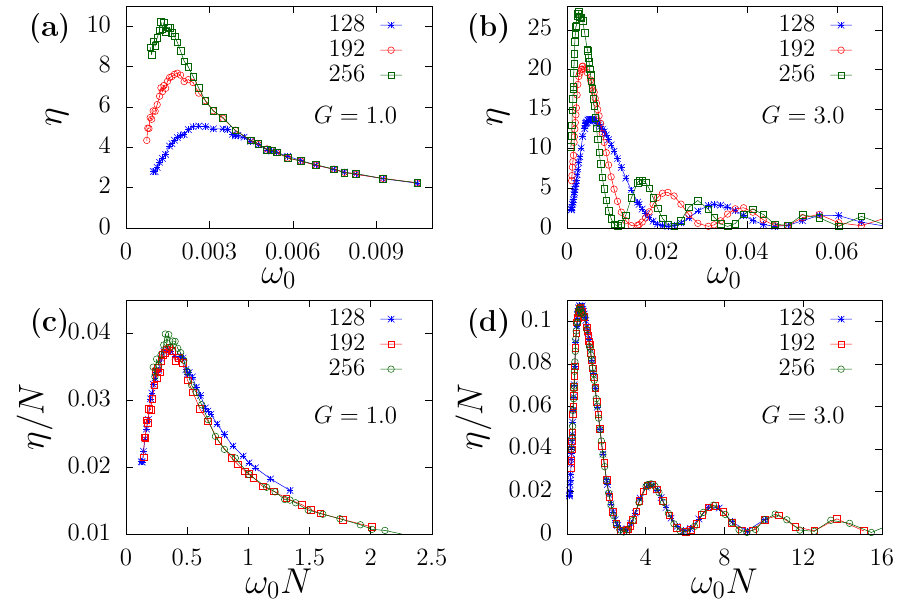}
	
		\caption{Variation of $\eta$ as a function of applied force
		frequency $\omega_0$ for three different chain lengths $N = 128,
		192,$ and $256$ at force amplitude (a) $G = 1.0$ and (b) $G =
		3.0$. When the scaled $\eta$/$N$ for various chain lengths are
		plotted with $\omega_0 N$ in accordance with Eq.~\ref{eq:4}, we
		get a nice data collapse for both (c) $G = 1.0$ and (d) $G =
		3.0$.}\label{OSN}

	\end{figure}
}

\newcommand{\figFive}{
	\begin{figure}[b]
    
		\includegraphics[width=0.5\textwidth]{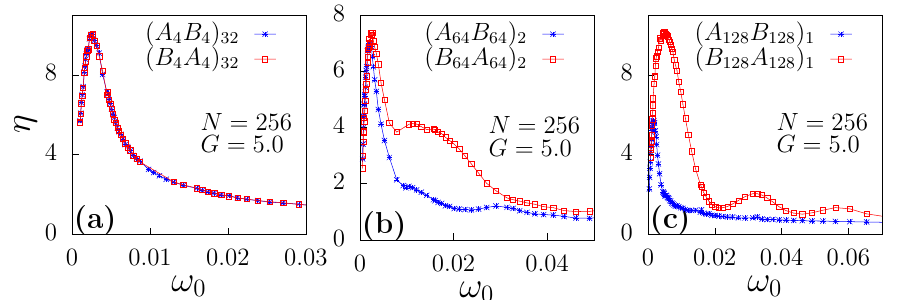}
	
		\caption{Output signal $\eta$ as a function of frequency
		$\omega_0$ of the driving force for a block copolymer DNA of
		length $N = 256$ at force amplitude $G = 5.0$ for the following
		sequences (a) $(A_{4}B_{4})_{32}$ and $(B_{4}A_{4})_{32}$, (b)
		$(A_{64}B_{64})_{2}$ and $(B_{64}A_{64})_{2}$, and (c)
		$(A_{128}B_{128})_{1}$ and $(B_{128}A_{128})_{1}$ at temperature
		$T = 4.0$. Note that in each of the plots the block sequences
		are opposite, implying that the end where the driving force is
		applied is the A base pair in one and the B base pair in the
		other. }\label{block}

	\end{figure}
}

\newcommand{\figTwo}{
	\begin{figure}[t]
    	\includegraphics[height=6cm]{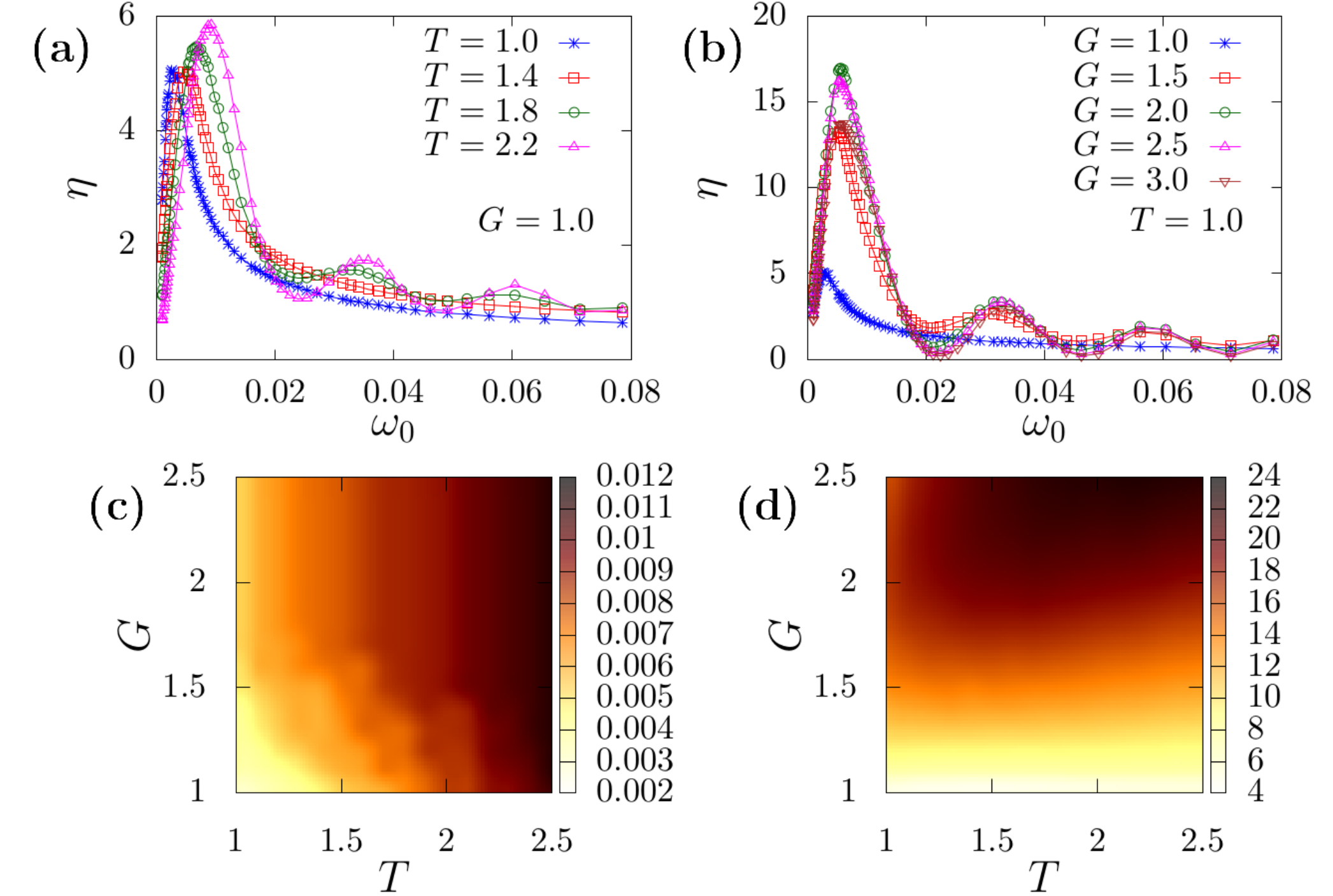}
	
		\caption{(a) Variation of $\eta$ as a function of frequency of
		the applied periodic force $\omega_0$ at a fixed force amplitude
		$G = 1.0$ and for four different temperatures $T$. (b) Variation
		of $\eta$ with $\omega_0$ at a fixed temperature $T = 1.0$ and
		for four different force amplitudes $G$.  Phase diagram in the
		$G-T$ plane plotted for (c) the primary peak frequency and (d)
		the height of the primary peak.}\label{phasediagram}

	\end{figure}
}

\newcommand{\figThree}{

	\begin{figure}[t]

		\centering

		\includegraphics[width=0.5\textwidth]{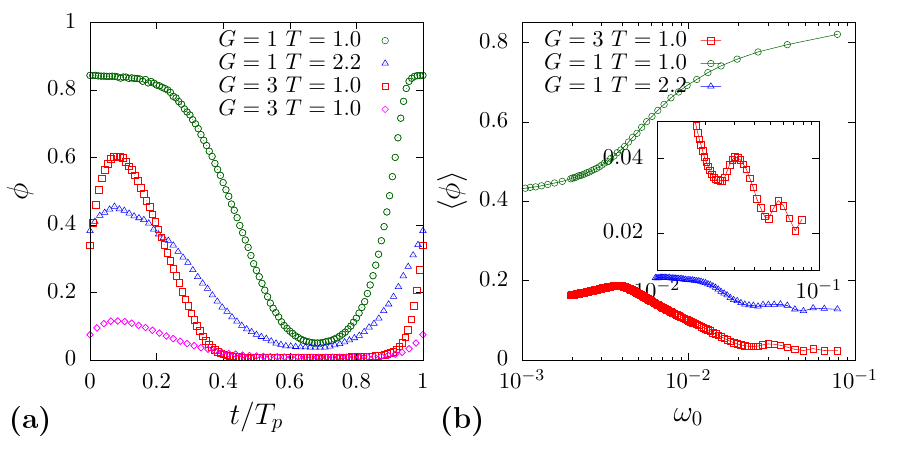}

		\caption{(a) Time dependence of the fraction of bound base pairs
		$\phi$ as a function of scaled time $t/T_p$ for various $G$, $T$
		and frequencies $\omega_0 = 2.62\times 10^{-3} \ (\circ) $,
		$9.23\times 10^{-3} \ (\triangle)$, $5.61 \times 10^{-3} \
		(\Box)$, and $9.23\times 10^{-3} \ (\Diamond)$. (b) Average value
		of the fraction of bound base pairs per cycle $\langle \phi
		\rangle$ as a function of frequency $\omega_0$ at various $G$
		and $T$ values. The inset shows secondary peaks for $G=3$ and
		$T=1$. }\label{fracBP}

	\end{figure}
}

\begin{abstract}

	We numerically study stochastic resonance in the unzipping of a model
	double-stranded DNA by a periodic force. We observe multiple peaks
	in stochastic resonance in the output signal as the driving force
	frequency is varied for different force amplitudes, temperature,
	chain length, and chain heterogeneity. Multiple peaks point to the
	existence of multiple stable and metastable states, which correspond
	to dynamical states of partially zipped and unzipped conformations
	and transitions between them. We quantify such transitions by
	looking at the time evolution of the fraction of bound base pairs.
	We obtain phase diagrams in the force amplitude--temperature plane
	both in the resonance frequency of the primary peak and the output
	signal at the peak value. We further obtain an excellent scaling
	behavior of the output signal for changing lengths of the DNA.
	Resonance behavior is also affected by chain heterogeneity as it
	depends strongly on which base pair the periodic forcing is applied. 

\end{abstract}

\maketitle

\textit{Introduction.} Stochastic Resonance
(SR)~\cite{benzi1981mechanism,gammaitoni1998stochastic,wellens2003stochastic}
has been studied in a wide variety of systems like the Brownian
particles in a double well
potential~\cite{simon1992escape,mahato1998some}, multithreshold
systems~\cite{gammaitoni1995stochastic}, neuron
models~\cite{wiesenfeld1994stochastic}, and quantum
systems~\cite{grifoni1996coherent,witthaut2009dissipation} to name a
few. First put forward by Benzi \textit{et
al.}~\cite{benzi1981mechanism}, it says that the response of a nonlinear
system is amplified at a certain frequency due to the noise within the
system (arising because of the probabilistic nature of the system) or by
the input noise. The frequency at which the response is maximum is
called the resonance frequency.  Recently Hayashi \textit{et
al.}~\cite{hayashi2012single} performed experimental studies of
unzipping of single DNA hairpins under the action of an oscillating
mechanical force applied with optical tweezers. When the force
oscillates around the average force needed to unfold the DNA, the
hopping kinetics between folded and unfolded states synchronizes with
the external frequency giving rise to SR.

Unzipping of a double-stranded DNA (dsDNA) is a first order phase
transition
\cite{bhattacharjee2000unzipping,danilowicz2004measurement,kapri2004complete,kalyan2015monte}.
Considering that the dependence of the free energy of the system on the
unzipping pathway coordinate has an asymmetric double well form; the
transition between a folded ground state and an unfolded metastable
state can be viewed as a barrier crossing
problem~\cite{volkov2009mechanism}. When an external periodic force is
applied, the free energy potential gets tilted asymmetrically up (down)
while raising (lowering) the energy barrier. SR activates when the
average waiting time between the two noise induced interwell transitions
of the system is comparable to the time period of the external periodic
force
\cite{gammaitoni1995stochastic,gammaitoni1998stochastic,gammaitoni1989periodically}.
During this process, the response of the system towards the external
force will be maximum. 

Studies on unzipping of a dsDNA by a periodic force have provided
valuable insight about the free energy landscapes that control the
folding
kinetics~\cite{hatch2007measurements,kumar2013statistical,kapri2014unzipping,pal2018dna,kalyan2019unzipping,yadav2021unzipping,kapri2021hysteresis}.
These studies have shown that by simply changing the frequency,
$\omega$, of an externally applied force, the dsDNA can be taken from a
zipped phase to an unzipped phase via a dynamical phase.  Hysteresis is
observed in this process due to the lag between response of the DNA to
the applied force. The area of this hysteresis loop increases with an
increase in the frequency up to a certain value $\omega^*$ and then
decreases on further increment in the frequency. Such a behavior is
reminiscent of SR, which we explore in this study and show that it can
be used as a measure of multiple dynamical transitions observed in the
system~\cite{huguet2009statistical,kapri2014unzipping,kalyan2019unzipping,yadav2021unzipping,kapri2021hysteresis}.

\figOne

We use Monte Carlo simulations to investigate SR in a model dsDNA driven
by a periodic force. We measure the SR by looking at the oscillations of
the separation of the end monomers of the strands where the periodic
force is applied. We calculate the output signal ($\eta$) defined as the
spectral density of the signal at an oscillation frequency and find that
this is an excellent quantifier of SR in our system. $\eta$ as a
function of the oscillation frequency at varying force amplitudes and
temperatures shows multiple resonance peaks which appear as signatures
of dynamical states of partially zipped and unzipped conformations and
transitions between them. The time evolution of the fraction of bound
base pairs and the average fraction of bound base pairs over a cycle
corroborates our findings. The peak frequency increases with increasing
chain lengths and shows excellent scaling behavior. We also looked at SR
signatures for a heterogeneous DNA where heterogeneity is introduced in
the form of repeated blocks with two different types of base pairs.
Remarkably, $\eta$ shows a sequence dependence: there is significant
difference in $\eta$ depending on which end is unzipped first.

\textit{Model.} In our model, the two strands of a homopolymer DNA are
represented by two directed self-avoiding walks on a
($d=1+1$)-dimensional square lattice.  The walks, which start from the
origin, are restricted to go towards the positive direction of the
diagonal axis ($z$ direction) without crossing each other [see
Fig.~\ref{fig:1}(a)]. The directional nature of the walks ensures
self-avoidance and the correct base pairing of the DNA strands. Two
complementary monomers of the strands are allowed to occupy the same
lattice site leading to a gain in energy $-\varepsilon$ ($\varepsilon
>0$). To see the response of the system to a periodic force, one end of
the DNA is anchored at the origin and a time-dependent periodic force
$g(t) = G |\sin (\omega_0 t)|$ is applied along the transverse direction
($x-$ direction) at the free end. $G$ denotes the amplitude and
$\omega_0 = \pi/T_p$ the frequency of the periodic force where $T_p$ is
the time period.

Monte Carlo (MC) simulations  of the model are performed
using the Metropolis algorithm. Individual strands undergo Rouse
dynamics involving local corner-flip or end-flip moves such that they do
not cross each other.  In the beginning of a Monte Carlo step, a random
monomer of a randomly chosen strand is flipped. The move is accepted if
it results in the overlapping of complimentary monomers giving rise to a
base pair between the two strands. If the move results in the unbinding
of a base pair, it is accepted with a Boltzmann probability $\xi =
\exp(- \varepsilon/ k_B T)$ for an interior monomer. If the monomer was
initially in unbound form and after the flip remains so, the move is
always accepted. If the move results in crossing of strands, it is
rejected. If the chosen monomer lies along a straight line, the
coordinate of the monomer and the change in energy would remain the same
after flipping to preserve the bond length. Therefore, such a move is
also accepted.  For the end monomer, we again calculate the change in
the energy ($\Delta E$) of the DNA configuration before and after the
flipping of the monomer. In this case, an additional energy due to the
pulling force, $g(t) \Delta x$, where $\Delta x$ is the change in the
distance after flipping the monomer, will also contribute. 
If the change in the energy
$\Delta E \leq 0$, the move is always accepted. If $\Delta E > 0$, the move
is accepted with the Boltzmann probability $\exp (-\Delta E/k_BT)$.
Time is measured in Monte Carlo steps,
with one step consisting of $2N$ flip attempts whether the move is
accepted or rejected. This is done to ensure that, on an average, in one
time increment every monomer of the two strands has a chance to flip.
Further, detailed balance is always satisfied throughout the simulation
except for moves involving the end monomers.  

\figTwo
 
Although the above model ignores finer details, such as bending rigidity
of the dsDNA, sequence heterogeneity, and stacking of base pairs, to
name a few, it has been found that the basic features, such as the first
order nature of the unzipping transition and the existence of a
re-entrant region allowing unzipping by decreasing temperature, are
preserved by this model
\cite{marenduzzo2001dynamical,marenduzzo2001phase}. For this model the
zero force melting takes place at a temperature $T_m =
\varepsilon/\left[k_B\ln(4/3)\right]$~\cite{kapri2012hysteresis}, where
$k_B$ is the Boltzmann constant. Throughout this paper, we have chosen
$\varepsilon = 1$ and $k_B = 1$.  We work with temperatures which are
lower than the melting temperature $T_m \approx 3.476$ to prevent
unzipping of DNA in the absence of a force. Most of the results
presented are for $N = 128$.

In our simulations, the distance between the end monomers of the two
strands, $x(t)$, as a function of time for various force amplitudes $G$
and frequencies $\omega_0$ are monitored. From Fig.~\ref{fig:1}(c), we
can see that $x(t)$ follows $g(t)$ but with a lag. From the time series
$x(t)$, we then obtain the power spectral density, $S(\omega)$, defined
as the Fourier transform of autocorrelation function of $x(t)$,
\begin{equation} 
	S(\omega)=\int_{-\infty}^{\infty}\langle x(t)x(t+\tau)\rangle
	\exp^{-i \omega \tau} d \tau .  
\end{equation} 
Here
$\langle {\bf \cdot} \rangle$ denotes the time average over the series.
From the power spectral density $S(\omega)$, we can define an output
signal $\eta$ as
\begin{equation} \label{OSeqn} 
	\eta = \lim_{\Delta \omega \to 0}\int_{\omega_0-\Delta
	\omega}^{\omega_0+\Delta \omega} S(\omega) d\omega.  
\end{equation}
$\eta$ serves as the resonance quantifier in our system. It is important
to note that, in the case of some bistable systems, $\eta$ does not show 
a clear peak at resonance frequency. In such cases, signal-to-noise (SNR) 
ratio is used to identify SR~\cite{stocks1995theoretical}. SNR is defined 
as $\textrm{SNR = $\eta$}/S(\omega_0)$.

We present results for the output signal at different frequencies
($\omega_0$) and amplitudes ($G$) of the force and also at different
temperatures ($T$). In our study, $S(\omega)$ is calculated from $x(t)$
using the TISEAN 3.0.1 software package~\cite{hegger1999practical}. The
average power spectral density is obtained by repeating the whole
process for 50 different initial states for each set of frequency,
amplitude, and temperature value.

\textit{Influence of force amplitude and temperature.} In
Fig.~\ref{fig:1}(b), we have plotted the time variation of the external
periodic force $g(t)$ for a DNA of length $N = 128$ at a given value of
$G$ and $\omega_0$ for three consecutive cycles. The corresponding time
variation of the extension $x(t)$ captures the folding and unfolding
dynamics of the DNA with a largely unzipped state at the maximum of
$x(t)$ and a zipped state when $x(t) \approx 0$ [Fig.~\ref{fig:1}(c)].
The Fourier transform of the stationary correlation function of $x(t)$
gives the power spectral density $S(\omega)$. As shown in
Fig.~\ref{fig:1}(d), $S(\omega)$ can be described as a superposition of
a background spectral density and delta peaks at frequencies which are
multiples of $\omega_0$~\cite{gammaitoni1998stochastic}. We then
calculate the output signal ($\eta$) using Eq.~(\ref{OSeqn}). 

In Fig.~\ref{phasediagram}(a), we have plotted $\eta$ as a function of
$\omega_0$ at a fixed force amplitude ($G = 1.0$) and for different
values of temperature. At lower temperatures, $\eta$ increases with
increasing frequency, reaching a peak value, and decreases as $\omega_0$
is increased further. For $T = 1.0, 1.4$, the force amplitude lies
slightly above the critical force $g_c$ needed to unzip the DNA and the
majority of the monomers of the DNA are in the zipped state. For very
low frequencies, the system is nearly in the equilibrium state and the
$\eta$ is low. On the other hand, at very high frequencies, the
oscillating force opens only a few of the base pairs and the resulting
$\eta$ is small. If we start at high frequencies and begin to lower the
frequency, the possibility of the DNA to relax to the oscillating force
increases, leading to more base pair openings. At a certain critical
frequency, we observe a peak in $\eta$ signaling SR
[Fig.~\ref{phasediagram}(a)].

As the temperature is raised ($T = 1.8,\ 2.2$), the situation is very
different. At these temperatures, we observe multiple smaller peaks in
$\eta$ at higher frequencies. The presence of multiple peaks in the
output signal is an indication of multiple stable and metastable states
in the system and transitions between them. At higher temperatures the
DNA is in an unzipped state and, with the oscillating force, the DNA
settles into metastable states of {\em partially} zipped conformations.
These partially zipped conformations occur near the base of the strands
away from the end where the force is applied. With changing frequency,
the system switches between these states resulting in secondary peaks in
OS.

While increasing the temperature leads to multiple peaks in OS, changing
temperatures can modify structural properties in an uncontrolled way by
changing thermodynamic stability and therefore are best avoided. A
better controlling parameter is the force amplitude. Indeed, in our
system, we observe the same behavior when the temperature is kept fixed,
while the force amplitude is varied [see Fig.~\ref{phasediagram}(b)].
The primary peak in the output signal again shifts to higher frequency
with increasing $G$. At higher values of $G$, secondary peaks of lower
strength in $\eta$ start to appear. At these force amplitudes, the
steady state of the system is one where the strands are completely
unzipped. At low frequencies, the system can relax and settle into a
zipped conformation giving rise to a primary peak in $\eta$. With
increasing frequencies, while it is difficult to settle into a
completely zipped state, metastable states of partially zipped
conformations are still possible. The occurrence of such states and the
transitions between them leads to the secondary peaks in $\eta$. This is
the first main result of this paper.

\figThree

To quantify our claim about metastable states with partially zipped
conformations, we looked at the time evolution of a fraction of bound
base pairs ($\phi$) averaged over $10^4$ cycles at a given frequency for
different values of force amplitudes ($G$) and temperatures ($T$) [see
Fig.~\ref{fracBP} (a)]. For all values of $G$ and $T$, $\phi$ oscillates
between (partially) zipped and (partially) unzipped states. For $G = 1$
and $T = 1$, we looked at the time evolution of $\phi$ at the frequency
where $\eta$ is maximum. $\phi$ starts from an almost completely zipped
state ($\phi \approx 0.8$) and with increasing $g(t)$ goes to an almost
unzipped configuration. At higher temperatures, $T = 2.2$, $\phi$
oscillates between a partially zipped $\phi \approx 0.45$ and a nearly
unzipped state. 

While the time evolution of $\phi$ provides a picture of the
transitions, it does not give a clear idea about the different peaks and
troughs in $\eta$. To do so, we looked at the average fraction of bound
base pairs ($\langle \phi \rangle$) over a cycle [see Fig.~\ref{fracBP}
(b)]. We plotted $\langle \phi \rangle$ with the frequency $\omega$ of
the periodic force at various values of $G$ and $T$. At $G = 1$ and $T =
1$, $\langle \phi \rangle$ increases monotonically with increasing
frequency indicating that the DNA is primarily in a zipped conformation.
Strikingly, at a higher temperature, $\langle \phi \rangle$ shows
oscillations with increasing frequency similar to that observed in
$\eta$. The same feature is observed at $G = 3, T = 1$. 

In Fig.~\ref{phasediagram}(c), we plot the phase diagram for the
position of the peak frequency in the $G - T$ plane. At fixed $G$, the
peak shifts to higher frequencies with increasing $T$. This shift is
however nonmonotonic with increasing $G$. At a higher $G$, the peak
frequencies are larger for the same temperature. In
Fig.~\ref{phasediagram}(d), we plot the phase diagram for the peak
height in the $G - T$ plane. While the peak height increases marginally
with increasing temperature for a fixed force amplitude, the variation
is largely monotonic with increasing $G$. 

\figFour
\figFive

\textit{Influence of changing DNA lengths.} In order to see what happens
when we vary the length of the DNA, we observed the response of the
system to periodic driving as $N$ is increased at a fixed temperature $T
= 1.0$.  In Fig.~\ref{OSN}(a), we plot the output signal as a function
of the driving frequency for three different chain lengths at $G = 1.0$.
There is only one resonance peak for these parameter values as observed
earlier. With increasing chain length, the peak height increases
significantly and the resonance peak is sharper. The peak also shifts to
lower frequencies with increasing length. In Fig.~\ref{OSN}(b), we plot
the output signal as a function of the driving frequency for three
different chain lengths at $G = 3.0$. As expected, there are multiple
secondary peaks for these parameter values. With increasing length, the
peak positions shift to lower frequencies and the peak height increases. 

The above observations indicate that, in the thermodynamics limit $N
\rightarrow \infty$, the resonance peak frequency $\omega^*(G)
\rightarrow 0$. This suggests a scaling of $\eta$ aa 
\begin{equation}
    \eta \sim N^{\delta}{\cal G}(\omega_0N^z),
    \label{eq:4}
\end{equation}
where $\delta$ and $z$ are critical exponents. We observe excellent data
collapse for both $G = 1.0$ and $G = 3.0$ with $\delta \approx 1$ and $z
\approx 1$ [Figs.~\ref{OSN}(c) and \ref{OSN}(d)]. This is the second
main result of this paper.

\textit{Influence of chain heterogeneity.} We further ask if chain
heterogeneity can lead to changes in the output signal. Heterogeneity in
the DNA chain is introduced in the form of repeated blocks, $-A_nB_n-$
or $-B_nA_n-$, where $2n$ is the block length, and $A$ and $B$ are
different types of base pairs with two- and three-hydrogen bonds,
respectively. The introduction of chain heterogeneity implies that the
end of the strands where the external periodic force is applied can be
either an $A$ or $B$ base pair. We consider the variation of $\eta$
as a function of frequency at two different force amplitudes $G$ for
different block lengths [$(A_4 B_4)_{32}$,  $(A_{64}B_{64})_{2}$,
$(A_{128} B_{128})_{1}$] as well as their opposite sequence [$(B_4
A_4)_{32}$,  $(B_{64}A_{64})_{2}$, $(B_{128} A_{128})_{1}$] for a chain
length $N=256$. 

The critical force is independent of the sequence of the DNA i.e., it
does not matter whether the DNA is unzipped from the end having base
pairing with three hydrogen bonds (stronger) or the base pairing with
two hydrogen bonds. For periodic forcing with $G = 5$, we find that, for
smaller block sizes like $(A_4 B_4)_{32}$, $\eta$ shows a single
resonance peak at lower frequency. For these block sizes, it again does
not matter which end is being unzipped first (see Fig.~\ref{block}(a)).
However, as the block sizes increase, $\eta$ starts showing additional
peaks at higher frequencies and there is a strong dependence on which
end is unzipped first. For $G = 5$, multiple peaks are observed for
sequences $-B_nA_n-$ [see Fig.~\ref{block}(b) and \ref{block}(c)]. With
periodic forcing, it is easier to unzip the polymer end if the bonding
is weaker. This is true where the pulling force is applied to A-type
base pairs, i.e., for the $-B_nA_n-$ sequences. The strands are therefore
separated easier and these separated strands can explore more
conformations. This difference in behavior of $\eta$ can serve as a
measure to differentiate between block copolymer sequences and is the
third main result of this paper. 

\textit{Discussions. } Motivated by experiments to probe how a DNA
molecule which exhibits bistability by hopping between folded and
unfolded conformations responds to an applied oscillating force, we
have carried out a numerical study of SR in a periodically driven DNA.
In line with experiments, we measured the power spectral density of the
extension of the strands of the DNA and observed the frequency
dependence of the output signal. The output signal quantifies SR in this
system. We looked at how the various parameters in our numerical study
such as the force amplitude, temperature, chain length, and heterogeneity
influence the resonance behavior of the system. 

A very interesting feature observed in our numerical study was the
existence of multiple peaks in $\eta$ at a given force amplitude.  The
presence of multiple peaks in SR points to the existence of multiple
stable and metastable states for a given force amplitude $G$. This is
equivalent to having multiple potential wells and particle hopping
between them. While such cases have been reported earlier in an array of
underdamped, nonlinear oscillators~\cite{lindner2001monostable} and in a
Fermi-Pasta-Ulam chain~\cite{miloshevich2009stochastic}, this is a
report of such a situation in a DNA chain.

While our numerical study exhibits some of the known features of SR in
periodically driven DNA chains and predicts several others including the
multiple peaks and a unique scaling behavior with increasing chain
lengths, it does not have the experimental features like trap stiffness
and DNA handles which have been shown to affect
SR~\cite{hayashi2012single}. Langevin dynamics simulations with detailed
set-up similar to experimental ones~\cite{kapri2021hysteresis} will be
able to give us a better understanding of the effect of these
parameters. In addition, we would like to explore the role of bond
stiffness and bending rigidity of the DNA in these systems. Further,
exploring other SR quantifiers like the average mechanical work per
period of the oscillation~\cite{mossa2009measurement} and extracting the
kinetic rates of hopping of molecules by measuring the frequency of
resonance will be the subject of future studies. 

\textit{Acknowledgments.} We acknowledge the use of the computing
facility at IISER Mohali. We thank an anonymous referee for valuable
suggestions.

\end{document}